\documentclass[fleqn]{article}
\usepackage{longtable}
\usepackage{graphicx}
\usepackage{color}

\begin {document}
\begin{flushleft}
{\LARGE
{\bf Effective  collision strengths for allowed transitions among the  $n \le$ 5   degenerate levels of
atomic hydrogen}
}\\

\vspace{1.5 cm}

{\bf {K  M  ~Aggarwal$^1$,  R Owada$^2$ and A Igarashi$^2$ }}\\ 

\vspace*{1.0cm}

$^1$Astrophysics Research Centre, School of Mathematics and Physics, Queen's University Belfast, Belfast BT7 1NN, Northern Ireland, UK\\ 
$^2$Department of Applied Physics, Faculty of Engineering, University of Miyazaki, Miyazaki 889-2192, Japan

\vspace*{0.5 cm} 

e-mail: K.Aggarwal@qub.ac.uk \\

\vspace*{1.50cm}

Received  4 June 2018\\
Accepted for publication 3 July 2018 \\
Published xx  July 2018 \\

\vspace*{1.5cm}

PACS Ref:  31.25 Jf, 32.70 Cs

\vspace*{1.0 cm}

\hrule

\end{flushleft}

\clearpage


\begin{abstract}

We report calculations of collision strengths and effective collision strengths for 26 allowed transitions among  the  $n \le$ 5   degenerate levels of atomic hydrogen for which the close-coupling (CC) and Born approximations have been used. Results are listed over a wide range of energies (up to 50~Ryd) and temperatures (up to 10$^7$~K), sufficient for applications over a variety of plasmas, including fusion. Similar results have also been calculated for deuterium,  but they negligibly differ with those of hydrogen.

\end{abstract}

\clearpage

\section{Introduction}

Hydrogen is the most abundant element in the universe and therefore, atomic data for its emission lines are very important for various studies of astrophysical plasmas. This element is equally important for fusion, because it is the main fuel for burning in a reactor, and therefore the importance of its data has further increased with the developing ITER project. Energies for its levels and radiative rates (A-values) for its transitions are fairly well known -- see for example, the compilations by  Kramida \cite{aek} and Wiese and Fuhr \cite{wf},  and the NIST (National Institute of Standards and Technology) website at {http://www.nist.gov/pml/data/asd.cfm}.  However, the corresponding information about the collisional data for electron impact excitation lacks completeness. Most of the data, experimental or theoretical, are limited in the range of energy or the number of transitions, as summarised  by Anderson et al.\, \cite{and1} and Benda and Houfek \cite{bh}, and further discussed below. 

The first major study of collisional data for H was performed by us (Aggarwal et al. \cite{kmh}), which included states with $n \le$ 5. The calculations were based on the close-coupling $R$-matrix method and reported results for both collision strengths ($\Omega$) and {\em effective} collision strengths ($\Upsilon$), obtained after integrating the $\Omega$ data over a {\em Maxwellian} distribution of electron velocities -- see a review by Henry \cite{henry}  for the general background about  electron atom/ion collisions.  A notable deficiency of this work was that higher ionisation channels, at energies above thresholds,  were not considered, and this leads to the overestimation of results, for some transitions. Therefore, Anderson et al. \cite{and1} included {\em pseudostates} in the expansion of wavefunctions in the $R$-matrix framework, and this allowed for the loss of electron flux into the continuum. They did not specifically list the $\Omega$ data but reported results for $\Upsilon$ for most (not all) transitions among the $n \le$ 5 states of H. Unfortunately, questions were soon raised about the accuracy of their work, particularly at higher temperatures. Therefore, they subsequently corrected their results in a later paper \cite{and2}. Nevertheless, doubts remain about the accuracy and reliability of their data, as discussed by Lavrov and Pipa \cite{lp} and W\"{u}nderlich et al.\, \cite{wdf}.

Recently, Benda and Houfek \cite{bh} have performed yet another calculation by employing a different approach, based on direct solution of the Schr{\"o}dinger equation in the B-spline basis. They also made some improvements over the earlier results and {\em concluded} an overall good agreement with the $\Omega$ data of Aggarwal et al.\, \cite{kmh}, although differences for a few transitions, particularly those from the ground 1s to higher excited states, are up to 12\%.  Although they presented their results for $\Omega$ only graphically, corresponding numerical data can be easily obtained, in a very fine energy mesh, from their website {http://utf.mff.cuni.cz/data/hex}. However,  they did not report the corresponding data for $\Upsilon$, which are required for the modelling or diagnostics of the plasmas, and neither can these  be calculated from their numerical data (except at very low temperatures), because of the limited energy range, below 1~Ryd. Therefore, practically the only reliable $\Upsilon$ data available for a larger number of transitions are those of Aggarwal et al.  Irrespective of the (in)accuracy of these (or other available) data, a major deficiency in the literature is that results for {\em fine structure} transitions, which are allowed  among the degenerate levels of  states, such as 3p~$^2$P$_{1/2,3/2}$ -- 3d~$^2$D$_{3/2,5/2}$ and 4d~$^2$D$_{3/2,5/2}$ -- 4f~$^2$F$_{5/2,7/2}$, are not yet available. This is because the degeneracy among these levels is practically zero and theoretically (very) very small -- see for example, the NIST website or present Table~1. For this reason, such transitions are often referred to as `elastic' and the calculations of  $\Omega$ for these are not only (very) sensitive to their energy differences ($\Delta \epsilon$), but are also very slow to converge --  see for example, figure~2 of Hamada et al.\, \cite{ham}  for a hydrogenic system Fe~XXVI, in which (depending on the energy) more than 10$^7$ partial waves were required to obtain the converged results. For the same reason, some of the fast atomic scattering codes, such as FAC (the flexible atomic code of Gu \cite{fac} and available at the website {https://www-amdis.iaea.org/FAC/}), cannot be confidently employed for such transitions, because discrepancies with the more accurate calculations can sometimes be large, as shown in figure~3 of Hamada et al. We will elaborate on this more later in Section~5.

The importance of $\Upsilon$ data for allowed fine structure transitions among degenerate levels has recently been reemphasised and demonstrated by Lawson et al.\, \cite{law} with respect to He~II, a hydrogenic ion and important for studying the fusion plasma. Since a code based on close-coupling method has already been developed by Igarashi et al.\, \cite{ai1, ai2} for calculating such data for hydrogenic ions,   and experience has been gained (Hamada et al.\, \cite{ham} and Aggarwal et al.\, \cite{kmhe}), we perform similar calculations for atomic hydrogen by suitably modifying the underlying theory. In addition, we perform calculations for deuterium (D) because it is also a part of the fusion fuel.

\section{Theory}  

For our calculations of $\Omega$ and $\Upsilon$ we use two methods described below.  

\subsection{Close-coupling method}  

 For calculating the electron-impact excitation among the fine-structure  levels of hydrogen atom, we  assume the contribution of the electron-exchange effect to be insignificant and hence neglect it. Furthermore, the validity of neglecting  the exchange effect has been discussed for the optically allowed transitions  of hydrogen-like targets   in our earlier work \cite{ai2}.  Similarly, in the close coupling expansion we only employ the physical states  with the same principal quantum number $n$, as in Igarashi et al. \cite{ai1}. 
 We also  treat the scattering electron as a spin-less  particle.  The total Hamiltonian of the system is defined as 
\begin{equation}
 \label{eq:totalH}
\everymath{\displaystyle}
\begin{array}{c}
   H= - \frac{1}{2} \nabla^2_r + h({\bf x}, s)   +  V({\bf r},{\bf x}), \\ 
    \rule{0in}{5ex}
     V({\bf r},{\bf x}) =  -\frac{1}{r} + \frac{1}{ | {\bf r} - {\bf x} |},\\
\end{array}
\end{equation}
 where ${\bf r}$ and ${\bf x}$ are the position vectors of
 the scattering 
 and the bound electron with respect to the nucleus, respectively,   $h$ is the Hamiltonian of hydrogen atom, and
 $s$ denotes the spin coordinates of the bound electron.

The atomic wave function of the fine-structure level may be approximated by
\begin{equation}
   \phi_{n \lambda j m_j} ({\bf x}, s) =
     R_{n \lambda}(x) \chi^{jm_j}_{\lambda 1/2} (\hat{x}, s), 
\end{equation}
 where $R_{n \lambda}$ is the non-relativistic radial function for
 the principal quantum number $n$ and the orbital angular momentum $\lambda$.
 The angular function is written by   
\begin{equation}
 \everymath{\displaystyle}
\begin{array}{ll}
      \chi^{j m_j}_{\lambda  1/2} ( \hat{x}, s) 
       =  &< \lambda \ \frac{1}{2} \ m_j-\frac{1}{2} \ \frac{1}{2} | j \ m_j>
           Y_{\lambda  m_j- \frac{1}{2} } (\hat{x}) \, \alpha(s)  \\
   \rule{0in}{5ex}  
     &  + < \lambda \ \frac{1}{2} \ m_j+\frac{1}{2} \ -\frac{1}{2} | j \ m_j>
           Y_{\lambda m_j+ \frac{1}{2} } (\hat{x}) \, \beta(s),
\end{array}
\end{equation}
 which is an eigen function of the angular  momentum $j$ (half integer) of the
  bound electron,
  its z-component $m_j$, and  the orbital angular momentum  $\lambda$. 
  The symbol $<l_1  l_2  m_1  m_2 | l_3  m_3>$ is the Clebsch-Gordan coefficient.
  The notations $\alpha$ and $\beta$  represent two-component spinors for spin {\it  up}
  and  {\it  down}, respectively. 
  An atomic basis is constructed 
  by  coupling the target wave function
  $ \phi_{n \lambda j m_j} $ with the angular function of the scattered
  electron as  
\begin{equation}
 \label{eq:basisa}
\everymath{\displaystyle}
\begin{array}{c}
  \psi^{JM_J\Pi}_{n \lambda  j l}( {\hat {\bf r}},  {\bf x}, s)
   = R_{n \lambda} (x)
    {\cal A}_{\lambda 1/2 (j)l}^{JM_J}(\hat{{\bf r}}, {\hat {\bf x}},s), \\
      \rule{0in}{5ex} 
  {\cal A}_{\lambda 1/2 (j) l}^{JM_J} = \sum_{m_j} < j  l  m_j  M_J-m_j|J  M_J>
       \chi^{j m_j}_{\lambda 1/2} (\hat{{\bf x}}, s) 
             Y_{l M_J-m_j} (\hat{{\bf r}}),
\end{array}
\end{equation}
 which is the eigen function of   the total angular momentum  $J$ (half integer),
 its z-component  $M_J$, and the parity $\Pi = (-1)^{\lambda+l} $.
 Using the atomic basis set 
 $ \{ \psi_\mu \} \equiv \{ \psi^{JM_J\Pi}_{n_\mu \lambda_\mu j_\mu l_\mu} \}$,
 the scattering wave function for the symmetry $\{ JM_J\Pi \}$ is expanded as 
\begin{equation}
  \label{eq:expansiona}
\everymath{\displaystyle}
\begin{array}{c}
  \Psi^{JM_J\Pi}({\bf r}, {\bf x}, s) = 
   \sum_\mu \frac{F_\mu(r)}{r}  \psi_\mu({\hat {\bf r}},{\bf x},s) .
\end{array}
\end{equation}   
 From the Schr\"odinger equation $\displaystyle (H-E) \Psi^{JM_J\Pi}=0$,
 we have a set of coupled equations for the radial function  $F_\mu$ as
\begin{equation}
 \label{eq:coupledeq}
\everymath{\displaystyle}
\begin{array}{c}
 \left( -\frac{1}{2} \left[ \frac{d^2}{dr^2} - \frac{l_\mu(l_\mu+1)}{r^2} \right] 
  - \frac{k_\mu^2}{2} \right) F_\mu(r) 
  +    \sum_\nu V_{\mu \nu}^\prime(r) F_\nu(r) =0,\\
\end{array}
\end{equation}
with  
\begin{equation}
\everymath{\displaystyle}
\begin{array}{c}
   k_\mu^2= 2(E - \epsilon_\mu), \\
    \rule{0in}{5ex}
   V_{\mu \nu}^\prime(r)= \sum_s
     \int d \hat{{\bf r}} \int d {\bf x} \   \psi_\mu^\ast({\hat {\bf r}},{\bf x},s) \, V({\bf r},{\bf x}) 
         \psi_\nu({\hat {\bf r}},{\bf x},s), 
\end{array}
\end{equation}
where $E$ is the total energy of the system and $\epsilon_\mu$ is 
 the atomic energy for  $\psi_\mu$.
 The coupled equations are solved under the
 boundary conditions
\begin{equation}
\everymath{\displaystyle}
\begin{array}{l}
     F_\mu^\nu(r=0)=0, \\
     F_\mu^\nu(r \to \infty) \sim  k_\mu^{-1/2} \left(  \delta_{\mu \nu}  
    \sin \Theta_\mu(r) +  \cos \Theta_\mu(r)  K_{\mu \nu}^{J \Pi} \right) ,\\    
  \rule{0in}{5ex}
 \Theta_\mu(r)=  k_\mu r - \frac{l_\mu \pi}{2}  
\end{array}
\end{equation}
  where 
  $ K_{\mu \nu}^{J \Pi}$ is the element of the K-matrix ${\bf K}^{J\Pi}$ for
     the symmetry $J$ and $\Pi$. 

  The transition matrix is given in terms of the K-matrix by
\begin{equation}
 \label{eq:tmatrix}
     {\bf T}^{J\Pi}= {\bf K}^{J\Pi} ({\bf 1} - i {\bf K}^{J\Pi})^{-1}.
\end{equation}
 The partial-wave cross section
 and the total cross section for transition from  atomic state $n \lambda  j$ to 
            $n^\prime \lambda^\prime j^\prime$ are given by 
\begin{equation} 
 \label{eq:pwcs}
\everymath{\displaystyle}
\begin{array}{c} 
 \sigma_ {n \lambda j ~n^\prime  \lambda^\prime j^\prime}^J
  =  \frac{4 \pi}{ (2j+1) k_{n\lambda j}^2 } \sum_{ll^\prime\Pi}  
       (2J+1) 
     | T_{n^\prime \lambda^\prime j^\prime l^\prime ~n \lambda j l}^{J\Pi} |^2,
\end{array}
\end{equation}
 and
\begin{equation}
 \label{eq:tcs}
\everymath{\displaystyle}
\begin{array}{c}
 \sigma_ {n \lambda j ~ n^\prime  \lambda^\prime j^\prime}
      =    \sum_J \sigma_ {n \lambda j ~ n^\prime  \lambda^\prime j^\prime}^J, 
  \end{array}
\end{equation}
  respectively.
 The collision strength (a dimensionless parameter) for transition between $n \lambda j$ and 
            $n^\prime  \lambda^\prime j^\prime$ is
  related to the cross section by  
\begin{equation}
     \label{eq:omega}
\everymath{\displaystyle}
\begin{array}{c}
 \Omega_ {n \lambda j ~  n^\prime  \lambda^\prime j^\prime}
  =   \frac{ k_{n \lambda j}^2 (2j+1)}{\pi}
       \sigma_{n \lambda j ~ n^\prime \lambda^\prime j^\prime}.
\end{array}
\end{equation}

\subsection{Born approximation}
In the Born approximation, 
  the scattering amplitude for the transition  $n \lambda j m_j$=${\rm i}\, m_j$ $\to$  
           $n \lambda^\prime j^\prime m_j^\prime$=${\rm i}^\prime \,  m_j^\prime$ is written as
\begin{equation}
\everymath{\displaystyle}
\begin{array}{l}
 f_{ {\rm i}m_j \to {\rm i}^\prime m_j^\prime} ({\bf q})
 = -\frac{1}{2\pi}  \left< e^{i {\bf k}^\prime \cdot {\bf r}}  \phi_{ {\rm i}^\prime  m_j^\prime} ({\bf x},s) \left| 
     \frac{1}{|{\bf r}-{\bf x}|} \right|   e^{i {\bf k} \cdot {\bf r}}  \phi_{ {\rm i} m_j} ({\bf x},s) \right>,    
\end{array}
\end{equation}
 where ${\bf q}= {\bf k} -{\bf k}^\prime$  is the momentum transfer.  
 The differential cross section   is given by
\begin{equation}
\everymath{\displaystyle}
\begin{array}{ccl}
 \frac{d \sigma_{{\rm ii}^\prime} }{d \Omega_{{\bf k}^\prime}} 
   &=& (2j+1)^{-1} \frac{k^\prime} {k} \sum_{m_j m_j^\prime}  |f_{ {\rm i}m_j \to {\rm i}^\prime m_j^\prime} ({\bf q})|^2 \\
   &=&  \frac{4k^\prime}{k  }     M_{{\rm ii}^\prime} (q)/q^2
\end{array}
\end{equation}
 with $ M_{{\rm ii}^\prime}(q) =  (2j+1)^{-1} \sum_{m_j m_j^\prime}  
  \left| \left<   \phi_{ {\rm i}^\prime  m_j^\prime} ({\bf x},s) \left| 
         {\rm e}^{ i {\bf q} \cdot {\bf x} } \right|  \phi_{ {\rm i} m_j} ({\bf x},s) \right>/q \right|^2$.  
 The integrated cross section is given by
\begin{equation}
 \everymath{\displaystyle}
\begin{array}{ccl}
 \label{eq:Bethe1}
  \sigma_{{\rm ii}^\prime} & = &
  \int \frac{d \sigma_{{\rm ii}^\prime}}{d \Omega_{{\bf k}^\prime}}   d \Omega_{{\bf k}^\prime} 
          = \frac{8\pi}{k^2} \int_{q_{\rm min}}^{q_{\rm max}} M_{{\rm ii}^\prime}(q) \frac{dq}{q} \\ \rule{0in}{4ex}
   &=& \frac{8\pi}{k^2} \int_{\log q_{\rm min}}^{\log q_{\rm max}} M_{{\rm ii}^\prime}(q) d (\log q)
\end{array}
\end{equation}
 with $q_{\rm min}=|k-k^\prime|$ and $q_{\rm max}=k+k^\prime$.
For the dipole allowed transitions, the integral over $q$ in (\ref{eq:Bethe1}) is determined mostly  by small $q$,
 and  $\sigma_{{\rm ii}^\prime}$ may be evaluated by the Bethe-Born approximation
 \begin{equation}
 \everymath{\displaystyle}
\begin{array}{c}
 \label{eq:Bethe2}
  \sigma_{{\rm ii}^\prime} \simeq   \frac{8\pi}{k^2} D_{{\rm ii}^\prime} (\log  {\bar q}_{\rm max} -\log q_{\rm min} )
\end{array}
\end{equation}
  with suitable ${\bar q}_{\rm max}$ value and  $ D_{{\rm ii}^\prime} =  (2j+1)^{-1} \sum_{m_j m_j^\prime}  
  \left| \left<   \phi_{ {\rm i}^\prime  m_j^\prime} ({\bf x},s) \left| 
         x \cos \theta_x  \right|  \phi_{ {\rm i} m_j} ({\bf x},s) \right> \right|^2$  -- see Section~2.3. 

 The Born cross section $\sigma_{{\rm ii}^\prime}$ in (\ref{eq:Bethe1}) can also be evaluated 
  by the partial-wave expansion   as  $\sigma_{{\rm ii}^\prime}= \sum \sigma_{{\rm ii}^\prime}^J$,
  and the partial-wave cross section is written as 
\begin{equation}
\everymath{\displaystyle}
\begin{array}{c}
 \sigma_ {n \lambda j ~ n^\prime \lambda^\prime j^\prime} ^J
  =  16 \pi  \sum_{ll^\prime}  
       (2J+1)  | I^J_{ll^\prime}|^2,\\
  \rule{0in}{5ex}
 I^J_{ l l^\prime} = \sum_s \int d{\bf r} d{\bf x} 
   \left[ j_{l^\prime} ( k^\prime r)   \psi_{n^\prime \lambda^\prime j^\prime l^\prime}^{J}(\hat{r},{\bf x},s) 
  \right]^\ast 
 \ V({\bf r}, {\bf x})
    \left[  j_l( k r)    \psi_{n \lambda j l}^{J}(\hat{r},{\bf x},s)   \right]  
\end{array}
\end{equation}
  where $j_l(kr)$ is the spherical Bessel function.

 The integral of the type
\begin{equation}
     \label{eq:CB2}
   {\tilde I}_{k_1  k_2}^l =  \int_0^\infty dr \,r^2\, j_l (k_1 r) j_{l+1} (k_2 r) \, \frac{1}{r^2} 
\end{equation}
     appears  for the optically allowed transitions,
 namely $|\lambda - \lambda^\prime|=1 $ and $|j-j^\prime| \le 1$. 
 When $ k_1 \simeq k_2 $,
 the integrand is of  quite a long-range and
 its values are important up to very large $l$.  
 Furthermore, the calculation  for ${\tilde I}^l_{k_1  k_2}$
  becomes numerically unstable for large  $l$,  
  but it can be well approximated (Alder et al.\, \cite{ak1, ak2}) as 
\begin{equation}
\everymath{\displaystyle}
\begin{array}{l}
 {\tilde I}^l_{k_1  k_2}  \simeq   2 /J_c 
    \left( K_1(a) - K_0(a) \right), \\  
   a=  \left( l + \frac{1}{2} \right) /J_c,  ~ ~ J_c= \sqrt{k_1 k_2} / |k_1 - k_2|
\end{array}
\end{equation} 
 using  modified Bessel functions. 
  Therefore,  $ {\tilde I}^l_{k_1  k_2} $ behaves  as 
\begin{equation}
 \label{eq:CB3}
\everymath{\displaystyle}
\begin{array}{c}
 {\tilde I}^l_{k_1  k_2} \propto 
 \Big \{  
     \begin{array}{ll}
      1/l   &  {\rm when} ~ l < J_c  \\  
     e^{- l/ J_c}   &  {\rm when} ~ l > J_c 
   \end{array}
\end{array}
\end{equation}
  due to the property of the modified Bessel functions.

\subsection{The choice of   ${\bar q}_{\rm max}$ for Bethe-Born approximation}
 Figure~1 shows the values of  $M_{{\rm ii}^\prime}(q)/D_{{\rm ii}^\prime}$ in (\ref{eq:Bethe1})
  of  (a) $np_{1/2} \to ns_{1/2}$  and (b) $ns_{1/2} \to np_{3/2}$   transitions  as functions of $n^2q$ ($n=2 \sim 5$).
  The value of    $M_{{\rm ii}^\prime}(q)$ approaches  that of $D_{{\rm ii}^\prime}$ for small $q$.
  The scaled curves for $n=2 \sim 5$ are in good accordance.
  They begin to decrease around $n^2 q \sim  0.1$ and are almost zero for $n^2q > 1$.
  Similar behaviours are seen for the other transitions. Therefore, we have set the value of ${\bar q}_{\rm max}$
  in (\ref{eq:Bethe2}) as
\begin{equation}
 \everymath{\displaystyle}
\begin{array}{c}
  {\bar q}_{\rm max}   =   \Biggl \{  
\begin{array}{ll}
      q_{\rm max}    &   q_{\rm max} \le  q_1  \\  
     (q_1 + q_{\rm max})/2   &  q_1 <  q_{\rm max} \le q_2 \\
      q_2                          &   q_{\rm max} > q_2
 \end{array}
\end{array}
\end{equation}
   with $q_1=  0.1/n^2$ and $q_2=  1/n^2$.

\section{Energy levels}

  As stated in Section~1, 
  the energy differences  between the fine-structure levels  within (any) $n$ are very small for H and D. 
 It is an important parameter for the optically allowed $nlj$--$nl^\prime j^\prime$ transitions.  
Energies for the levels of H and D have been taken from the NIST website 
{https://physics.nist.gov/PhysRefData/HDEL/difftransfreq.html}. 
   These are also listed in Table~1a for a ready reference, where
   the energy  of  an $nlj$  is presented as difference from the
  $n$p$_{1/2}$ level, which is the lowest  within an $n$ manifold.   
  Though the binding energy of D($nlj$) is  slightly  larger than that of H($nlj$), due to the isotope shift,
  the energies listed in Table~1a  for H are quite similar to those of D,  and are specifically listed in Table~1b, in increasing order.   
  This table also provides level indices for future references. 
  It may be noted that   the energy differences between  $nlj$ and $nl^\prime j^\prime$  levels are also similar,
   and  approximately  scale as $1/n^3$ for both H and D.  
 
\section{Partial cross sections and collision strengths}

As examples,
  we show in Figures  2--5 the variation of  $J \sigma^J$ 
 with $J$ for two excitation transitions   within
 $n$ = 2 ($i \to i^\prime$ = 2p$_{1/2}$ $\to$ 2s$_{1/2}$ and  2s$_{1/2}$ $\to$ 2p$_{3/2}$),
 3 (3s$_{1/2}$ $\to$ 3p$_{3/2}$ and  3d$_{3/2}$ $\to$ 3p$_{3/2}$),
 4 (4f$_{5/2}$ $\to$ 4d$_{5/2}$ and  4d$_{5/2}$  $\to$ 4f$_{7/2}$), and 
 5 (5d$_{5/2}$ $\to$ 5f$_{7/2}$ and  5g$_{7/2}$ $\to$ 5f$_{7/2}$), respectively,
 and at four incident energies of $E_i=k_i^2/2$ = 4$\times$10$^{-4}$, 1$\times$10$^{-2}$, 0.2, and 4 Ryd, which cover a wide range. 
 Two sets of results are shown in these figures, i.e. Born (broken curves) and CC+Born (continuous curves).  It is clear from these figures that the Born cross sections are overestimated  at most energies  (see also Figure~6),    particularly for $J <$ 100,  but above it there are (practically) no differences between the two sets of results. For this reason the CC results are only for $J <$ 100 and beyond these Born alone are used.
 The curves for $J \sigma^J$generally increase   steeply    with J at small values, then show plateaus at intermediate  ones, 
   and finally decrease exponentially for $J >  J_c=\sqrt{k_i k_{i^\prime}}/|k_i-k_{i^\prime}|$, as equation (\ref{eq:CB3}) indicates. 
  Note that $J_c \simeq E_i/(2 \Delta \epsilon) $ for $E_i \gg  \Delta \epsilon$, 
 where $ \Delta \epsilon = E_i -E_{i^\prime}$ is the excitation energy.   
As may be seen, particularly from Figure~5, that over 10$^{11}$ partial waves are required before $J\sigma^J$ starts decreasing.
 The  values of $J\sigma^J$ at the plateau regions decrease as $1/E_i$ with $E_i$ in Figures~2--5.

  In Table~2 we list our results for $\Omega$ for all 26 allowed {\it fine structure} transitions among the degenerate 
   levels for $n=2 \sim 5$ of H (given in Table~1b) in an energy range below 50~Ryd.  
Additionally,    in  Figure~6 we show the variation of  $\Omega$ with energy for the optically allowed transitions within the $n=4$ manifold, as  examples.
 The differences in the results obtained in the CC+Born and Born approximations gradually decrease with increasing energies, and  almost disappear above $\sim$10~Ryd. A similar trend has been observed for other transitions as well. Finally, 
 the  Bethe-Born calculations in equation (\ref{eq:Bethe2}), for both $\sigma$ and $\Omega$,  estimated by the three parameters  $D_{ii^\prime}$, $q_{min}$ and $q_{max}$,  reproduce well the corresponding results in the Born approximation. 
 
 Our $\Omega$ results, in both the CC+Born and Born approximations, for D are also included in Figure~6. However, these  are very close to the
    corresponding results for  H, and hence the differences between the two are negligible. 

\section{Effective collision strengths}

 The values of $\Omega$ listed in Table~2 are  averaged over the Maxwellian distribution of electron velocities to  obtain the {\em effective} collision strengths $\Upsilon$  as follows: 

\begin{equation}
  \label{eq:upsilon}
\Upsilon(T_e) = \int_{0}^{\infty} {\Omega}(E_{i^\prime}(E_i)) \, {\rm exp}(-E_{i^\prime}/kT_e) \,d(E_{i^\prime}/{kT_e}),
\end{equation}
where $k$ is Boltzmann constant, $T_e$ is the electron temperature in K, and $E_{i^\prime}$ 
 is the electron energy with respect to the  final (excited) state/level. This value of $\Upsilon$ is
related to the excitation $q(i,i^\prime)$ and de-excitation $q(i^\prime,i)$ rates as follows:

\begin{equation}
q(i,i^\prime) = \frac{8.63 \times 10^{-6}}{{\omega_i}{T_e^{1/2}}} \Upsilon \, {\rm exp}(- \Delta \epsilon/{kT_e}) \hspace*{1.0 cm}{\rm cm^3s^{-1}}
\end{equation}
and
\begin{equation}
q(i^\prime,i) = \frac{8.63 \times 10^{-6}}{{\omega_{i^\prime}}{T_e^{1/2}}} \Upsilon \hspace*{1.0 cm}{\rm cm^3 s^{-1}},
\end{equation}
where $\omega_i$ and $\omega_{i^\prime}$ are the statistical weights of the initial ($i$) and final ($i^\prime$) states, respectively, and $ \Delta \epsilon =E_i-E_{i^\prime}$ is the excitation energy.  We have used the $\Omega(E_i)$ values in the CC+Born  approximation for $E_i \le 20$~Ryd in equation (\ref{eq:upsilon}),   but in Born (equation (\ref{eq:Bethe1})) alone above it, and up to an energy of 1000~Ryd. 
 Results for these rates are required in the modeling and diagnostics of plasmas. The calculated values of $\Upsilon$ are listed in Table~3 for all 26 transitions and at a wide temperature range, up to 10$^7$~K, suitable for applications in a variety of plasmas, including fusion. Furthermore, the variation of $\Upsilon$ with $T_e$ is rather smooth, 
as between 10$^3$ and 10$^7$~K it varies by a maximum factor of 2.2 for a few transitions, and much less for most. Therefore, values of $\Upsilon$ at any desired $T_e$ within this range can be easily interpolated without any loss of accuracy.

In the absence of any other similar results in the literature, either for $\Omega$ or $\Upsilon$, it is difficult to assess the accuracy of  our  calculated data, particularly when the calculations are very sensitive to $\Delta \epsilon$, as noted in Section~1.  In the past, for several ions we have performed calculations with FAC to make some estimation of the accuracy of data -- see for example, Hamada et al.\, \cite{ham} for O~VIII and Ni~XXVIII and Aggarwal et al.\, \cite{kmhe} for He~II. However, similar calculations performed for H have been highly unsatisfactory, because for some transitions, such as 5--6 (3p$_{1/2}$--3s$_{1/2}$), 8--9 (3p$_{3/2}$--3d$_{5/2}$) and 10--12 (4p$_{1/2}$--4d$_{3/2}$), values of $\Omega$ decrease with increasing energy, whereas for others, such as 7--8 (3d$_{3/2}$--3p$_{3/2}$) and 10--11 (4p$_{1/2}$--4s$_{1/2}$), they suddenly increase by up to six orders of magnitude. Therefore, these calculations cannot be relied upon for comparison or accuracy assessment, and it confirms yet again that although FAC is very efficient for generating large amount of atomic data with (normally) some measure of accuracy, it is not designed to produce sophisticated results for higher accuracy. Nevertheless,  in the near future we plan to make a detailed analysis of population modelling for fusion (tokamak) plasmas, on a similar line as recently done for He~II (Lawson et al.\, \cite{law}), and that may give some idea of the accuracy of the reported data. 

Finally, we have (mostly) presented results for the transitions of H, but calculations have also been performed for {\em all} transitions of D. However, there are negligible differences ($<$ 1\%) between the two sets of results, for both $\Omega$ and $\Upsilon$,   because  $\Delta \epsilon$ among the levels  of H and D differ by  no more than $\sim$ 0.15\%. Therefore, we can confidently state that the same results as listed in Tables~2 and 3 can be reliably  applied for both H and D.

\section{Conclusions}

In this paper we have presented results for 26 transitions of H which are {\em allowed} within the degenerate levels of states with $n \le$ 5. Results have been listed for both $\Omega$ and $\Upsilon$ over a wide energy/temperature range, and will not only be helpful for the modelling of plasmas, but also for future comparisons, because no such data presently exist in the literature. The listed results also complement our earlier data (Aggarwal et al.\, \cite{kmh}) and are required for considering a complete plasma model for fusion studies. Additionally, parallel calculations have also been performed for D, but the results are insignificantly different from those for H. This is an important conclusion for future studies.

\section*{Acknowledgment}
We thank Professor Francis Keenan and Dr  Kerry Lawson for initiating our interest in this work and making us realise its importance.

\clearpage
 \setcounter{figure}{0}
 \begin{figure*}
\includegraphics[width=\columnwidth]{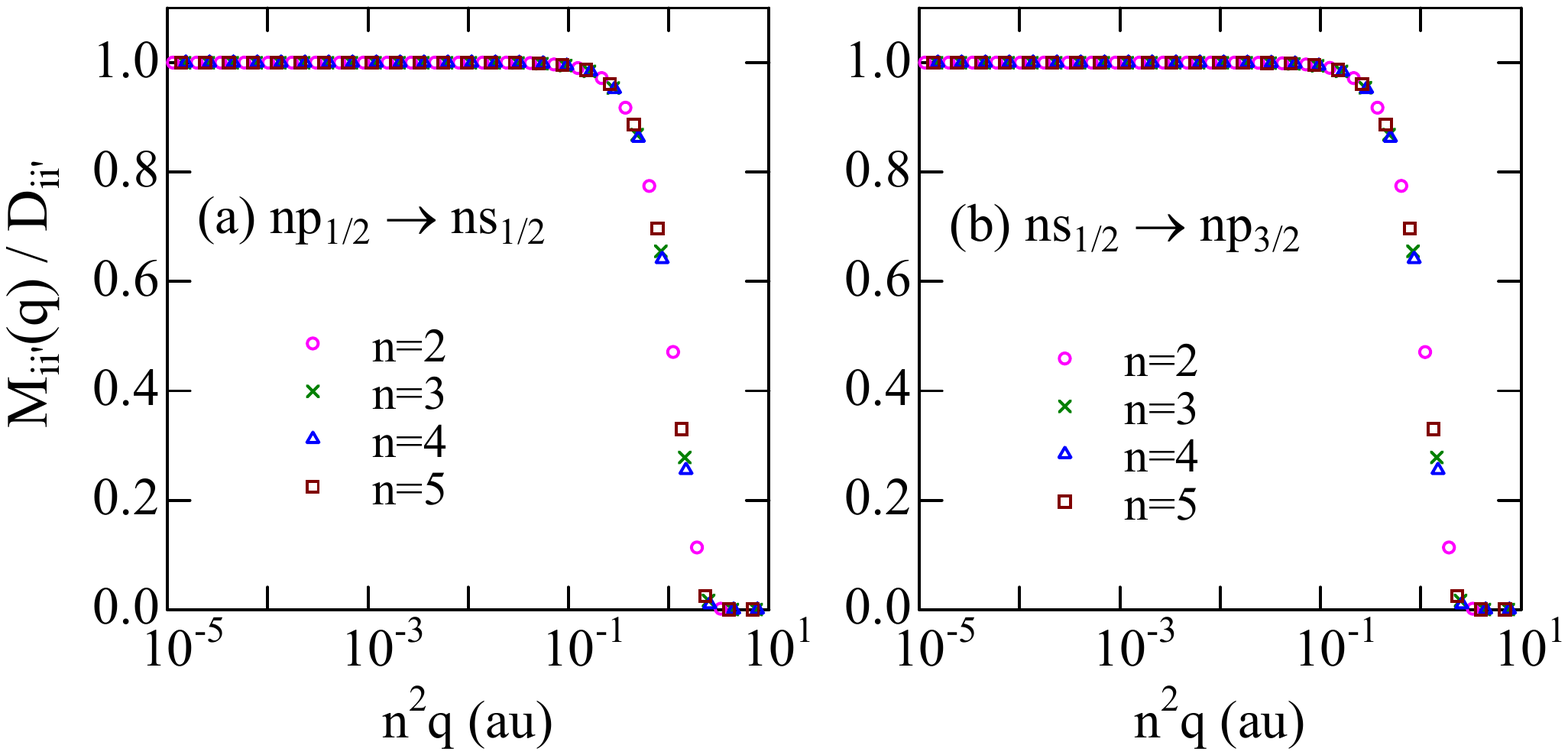}
 \vspace{-3.5cm}
 \caption{Values of  $M_{{\rm ii}^\prime}(q)/D_{{\rm ii}^\prime}$ for the (a) $np_{1/2} \to ns_{1/2}$  and (b) $ns_{1/2} \to np_{3/2}$ transitions as functions of $n^2q$ ($n=2 \sim 5$).}
 \end{figure*}

 \setcounter{figure}{1}
 \begin{figure*}
\includegraphics[width=\columnwidth]{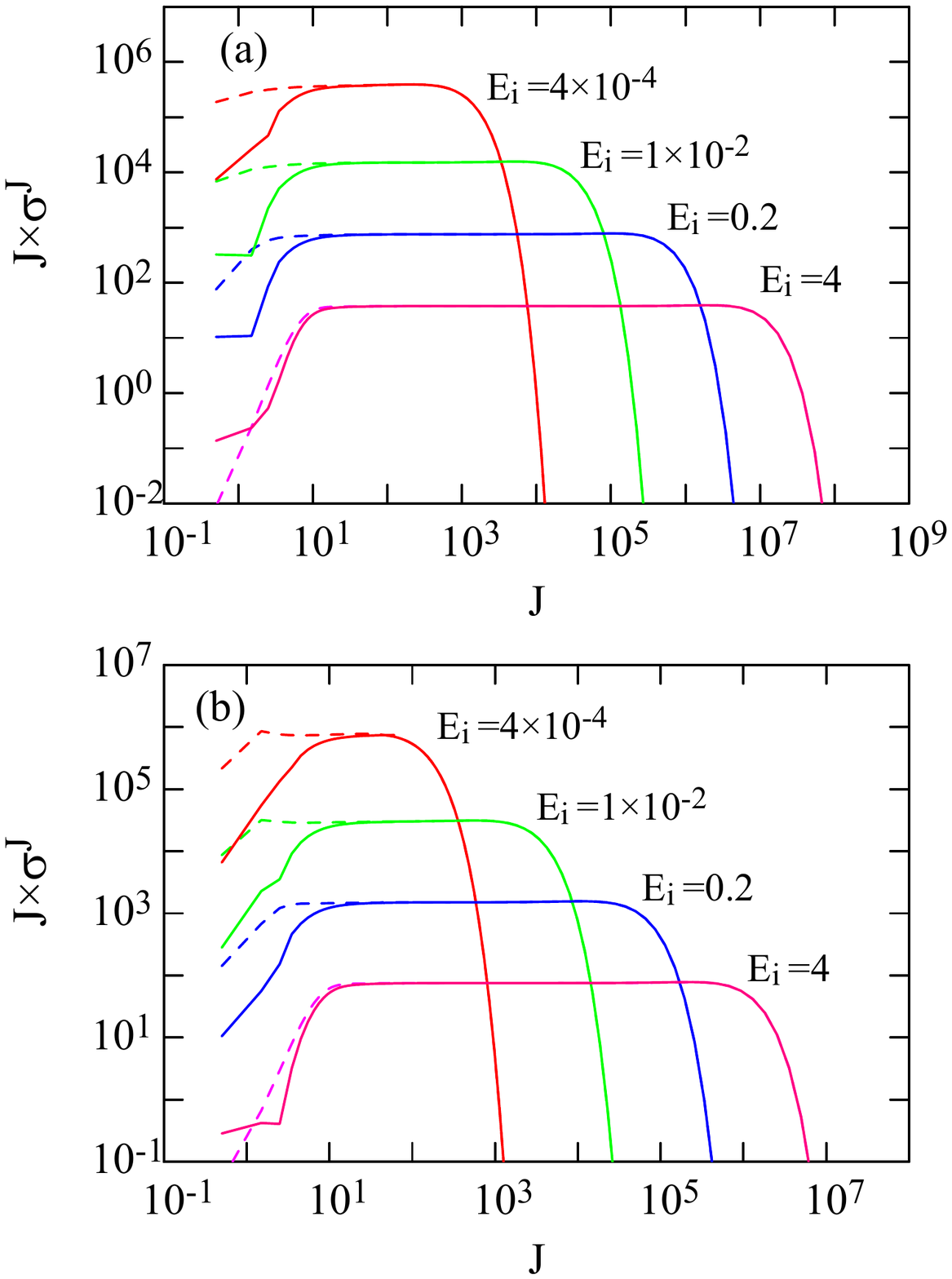}
 \vspace{-4.5cm}
 \caption{Variation of $J\times\sigma_J$  with partial waves $J$ for (a) 2p$_{1/2}$--2s$_{1/2}$ and (b) 2s$_{1/2}$--2p$_{3/2}$ transitions. Broken curves: Born and continuous curves: CC+Born results.}
 \end{figure*}

 \setcounter{figure}{2}
 \begin{figure*}
\includegraphics[width=\columnwidth]{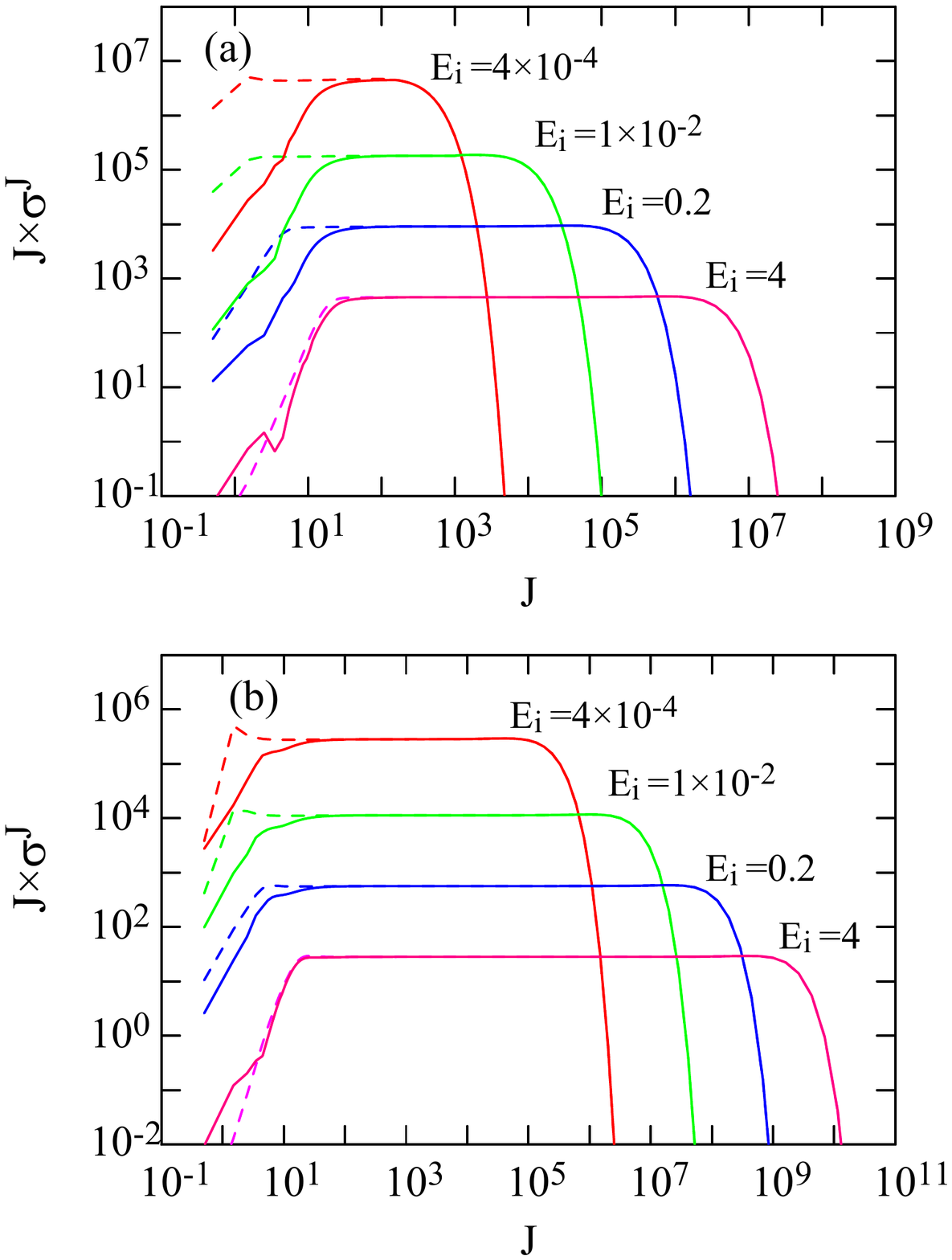}
 \vspace{-4.5cm}
 \caption{Variation of $J\times\sigma_J$ with partial waves $J$ for (a) 3s$_{1/2}$--3p$_{3/2}$ and (b) 3d$_{3/2}$--3p$_{3/2}$ transitions. Broken curves: Born and continuous curves: CC+Born results.}
 \end{figure*}
 \setcounter{figure}{3}
 \begin{figure*}
\includegraphics[width=\columnwidth]{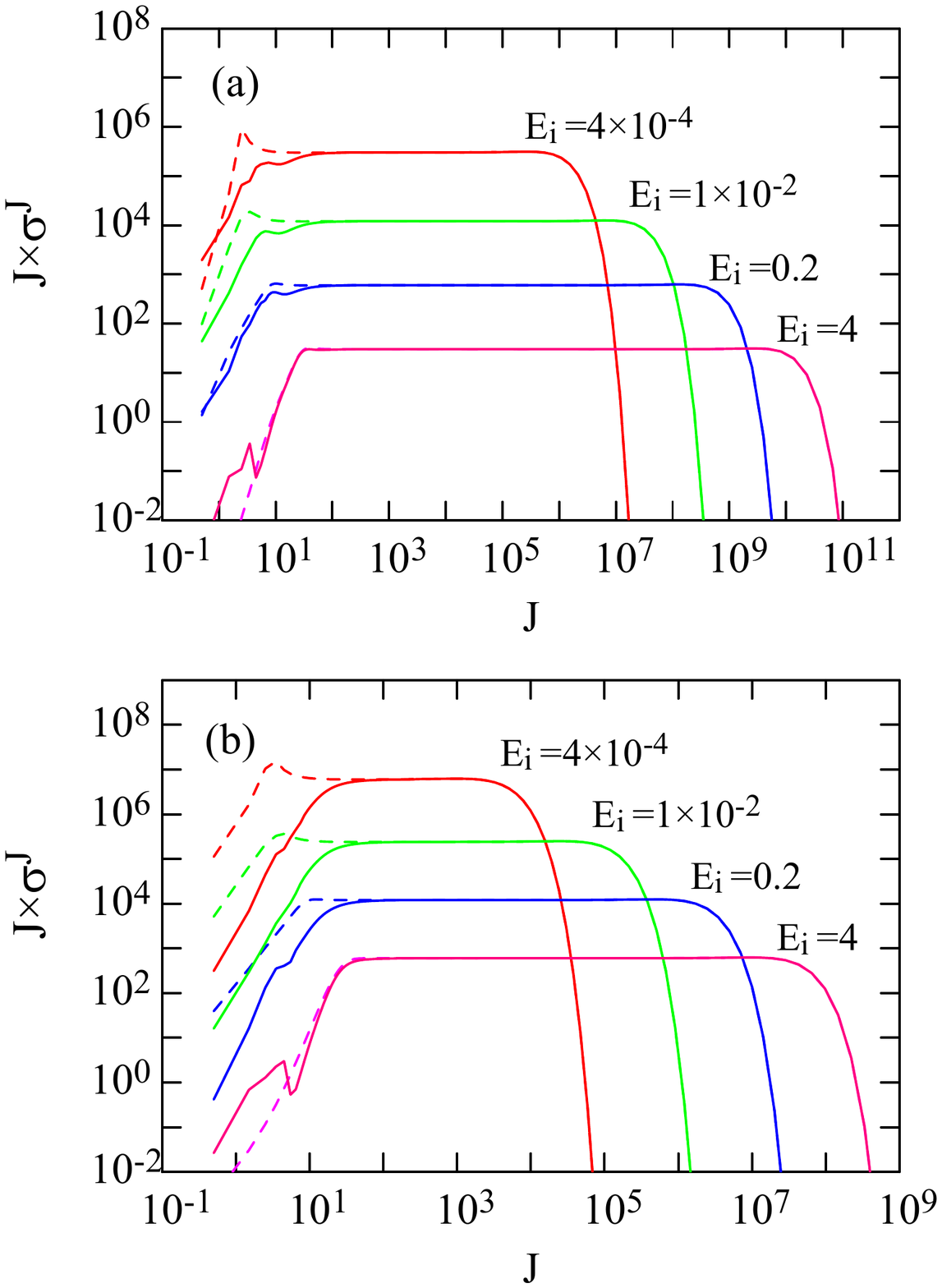}
 \vspace{-4.5cm}
 \caption{Variation of $J\times\sigma_J$  with partial waves $J$ for (a) 4f$_{5/2}$--4d$_{5/2}$ and (b) 4d$_{5/2}$--4f$_{7/2}$ transitions. Broken curves: Born and continuous curves: CC+Born results.}
 \end{figure*}

 \setcounter{figure}{4}
 \begin{figure*}
\includegraphics[width=\columnwidth]{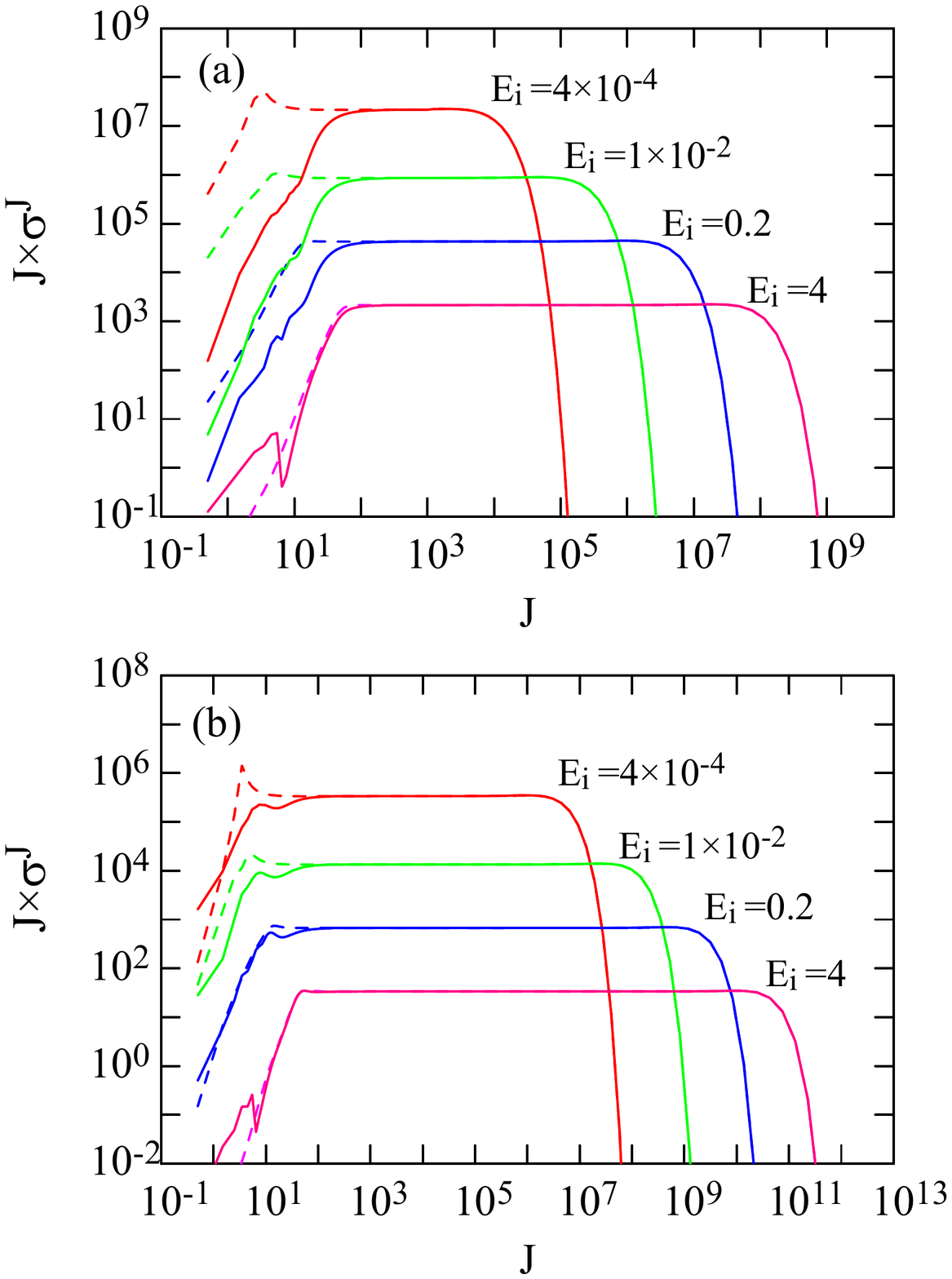}
 \vspace{-4.5cm}
 \caption{Variation of $J\times\sigma_J$  with partial waves $J$ for (a) 5d$_{5/2}$--5f$_{7/2}$ and (b) 5g$_{7/2}$--5f$_{7/2}$ transitions. Broken curves: Born and continuous curves: CC+Born results.}
 \end{figure*}

\setcounter{figure}{5}
 \begin{figure*}
\includegraphics[width=\columnwidth]{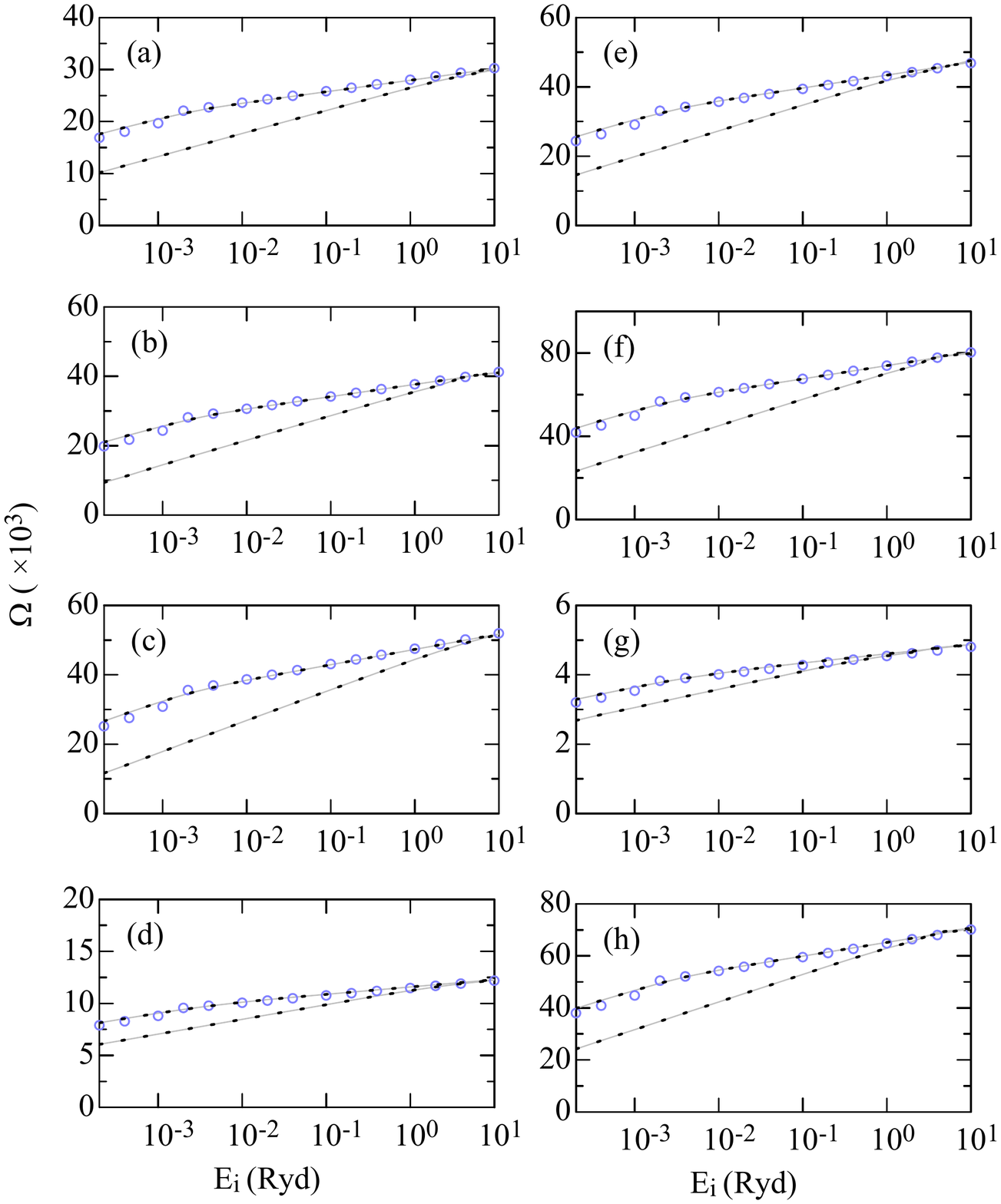}
 \vspace{-3.5cm}
 \caption{Variation of $\Omega$  with  energy $E_i$ for the  transitions within $n$ = 4.
(a) 4p$_{1/2}$--4s$_{1/2}$, (b) 4p$_{1/2}$--4d$_{3/2}$, (c) 4s$_{1/2}$--4p$_{3/2}$, (d) 4d$_{3/2}$--4p$_{3/2}$, 
(e) 4d$_{3/2}$--4f$_{5/2}$, (f) 4p$_{3/2}$--4d$_{5/2}$, (g) 4f$_{5/2}$--4d$_{5/2}$  and (h) 4d$_{5/2}$--4f$_{7/2}$. 
Upper and lower curves are for Born and CC+Born collision strengths for H (continuous curves) and D (dots).
Circles on upper curves are Bethe-Born results for H.}
 \end{figure*}

\begin{table*} 
\caption{a. Energies  (Ryd) for the levels of atomic hydrogen and deuterium with respect to the $n$p$_{1/2}$ level. $a{\pm}b \equiv a{\times}$10$^{{\pm}b}$.} 
\begin{tabular}{rlcccrrrrrrr} \hline
$n$  &  Level        &   H & D      \\
\hline
 2 & 2p$_{1/2}$ &  0.000000+0   &  0.000000+0 \\
   & 2s$_{1/2}$ &   3.215489$-$7  & 3.219711$-$7   \\      
   & 2p$_{3/2}$ &   3.334219$-$6  & 3.335129$-$6   \\  
  \noalign{\smallskip}
3  & 3p$_{1/2}$ &   0.000000+0 & 0.000000+0 \\
   & 3s$_{1/2}$ &   9.571226$-$8  & 9.583735$-$8  \\
   & 3d$_{3/2}$ &   9.862969$-$7  & 9.865650$-$7 \\
   & 3p$_{3/2}$ &   9.879176$-$7  & 9.881873$-$7  \\ 
   & 3d$_{5/2}$ &   1.315595$-$6  & 1.315953$-$6  \\
         \noalign{\smallskip}
 4 & 4p$_{1/2}$ &  0.000000+0  &    0.00000+0  \\
   & 4s$_{1/2}$ &   4.045105$-$8  &  4.050383$-$8  \\
   & 4d$_{3/2}$ &  4.160819$-$7   & 4.161950$-$7   \\
   & 4p$_{3/2}$ &  4.167772$-$7   & 4.168910$-$7   \\
   & 4f$_{5/2}$ &   5.547586$-$7   & 5.549094$-$7  \\
   & 4d$_{5/2}$ &  5.550046$-$7   & 5.551556$-$7   \\
   & 4f$_{7/2}$ &  6.242196$-$7   & 6.243893$-$7   \\
         \noalign{\smallskip}
 5 & 5p$_{1/2}$ & 0.000000+0  &   0.000000+0  \\
   & 5s$_{1/2}$ &  2.072911$-$8  &  2.075613$-$8   \\
   & 5d$_{3/2}$ &  2.130304$-$7  & 2.130883$-$7    \\
   & 5p$_{3/2}$ &  2.133897$-$7  & 2.134479$-$7    \\
   & 5f$_{5/2}$ &  2.840319$-$7   & 2.841091$-$7   \\
   & 5d$_{5/2}$ &  2.841589$-$7  & 2.842362$-$7    \\
   & 5g$_{7/2}$ &  3.195295$-$7  & 3.196164$-$7    \\
   & 5f$_{7/2}$ &  3.195959$-$7  & 3.196829$-$7    \\
   & 5g$_{9/2}$ &  3.408679$-$7  &  3.409606$-$7   \\
\hline 													     
\end{tabular}   
\end{table*}

\setcounter{table}{0} 
\begin{table*} 
\caption{b. Energy levels  (Ryd) of atomic hydrogen and deuterium. } 
\begin{tabular}{rlcccrrrrrrr} \hline
Index  &  Level        &   H & D      \\
\hline
  1   &   1s$_{1/2}$   &   0.000~000~000~000  &  0.000~000~000~000  \\
  2   &   2p$_{1/2}$   &   0.749~598~426~021  &  0.749~802~385~029  \\                     
  3   &   2s$_{1/2}$   &   0.749~598~747~570  &  0.749~802~707~000  \\
  4   &   2p$_{3/2}$   &   0.749~601~760~240  &  0.749~805~720~158  \\
  5   &   3p$_{1/2}$   &   0.888~414~397~520  &  0.888~656~128~310  \\
  6   &   3s$_{1/2}$   &   0.888~414~493~232  &  0.888~656~224~147  \\
  7   &   3d$_{3/2}$   &   0.888~415~383~817  &  0.888~657~114~875  \\
  8   &   3p$_{3/2}$   &   0.888~415~385~438  &  0.888~657~116~497  \\
  9   &   3d$_{5/2}$   &   0.888~415~713~115  &  0.888~657~444~263  \\
 10   &   4p$_{1/2}$   &  0.936~999~852~243   &  0.937~254~802~720 \\
 11   &   4s$_{1/2}$   &   0.936~999~892~694  &  0.937~254~843~224 \\
 12   &   4d$_{3/2}$   &   0.937~000~268~325  &  0.937~255~218~915 \\
 13   &   4p$_{3/2}$   &   0.937~000~269~020  &  0.937~255~219~611 \\
 14   &   4f$_{5/2}$   &   0.937~000~407~002  &   0.937~255~357~629 \\ 
 15   &   4d$_{5/2}$   &   0.937~000~407~248  &  0.937~255~357~876 \\
 16   &   4f$_{7/2}$   &   0.937~000~476~463  &  0.937~255~427~109 \\
 17   &   5p$_{1/2}$   &   0.959~487~919~100  &  0.959~748~987~500 \\ 
 18   &   5s$_{1/2}$   &   0.959~487~939~829  &  0.959~749~008~256 \\
 19   &   5d$_{3/2}$   &    0.959~488~132~130  & 0.959~749~200~588 \\
 20   &   5p$_{3/2}$   &   0.959~488~132~490  &  0.959~749~200~948 \\
 21   &   5f$_{5/2}$   &    0.959~488~203~132  &  0.959~749~271~609 \\
 22   &   5d$_{5/2}$   &   0.959~488~203~259  &  0.959~749~271~736 \\
 23   &   5g$_{7/2}$   &   0.959~488~238~630  &  0.959~749~307~116 \\
 24   &   5f$_{7/2}$   &   0.959~488~238~696  &  0.959~749~307~183 \\
 25   &   5g$_{9/2}$   &   0.959~488~259~968  &  0.959~749~328~461 \\
\hline 													     
\end{tabular}   
\begin{flushleft}
\end{flushleft}
\end{table*}

\begin{table*}
\caption{Collision strengths for transitions of atomic hydrogen as a function of energy. $a{\pm}b \equiv a{\times}$10$^{{\pm}b}$.}
\begin{tabular}{ccllllllllll} \hline
  $E_i$ (Ryd) & \multicolumn{8}{c}{Transition}         \\   \cline{2-10}
      I & 2: 2p$_{1/2}$ &   3: 2s$_{1/2}$  &  5: 3p$_{1/2}$   &  5: 3p$_{1/2}$    &  6: 3s$_{1/2}$  &   7: 3d$_{3/2}$  &   8: 3p$_{3/2}$  &   10: 4p$_{1/2}$ &  10: 4p$_{1/2}$  \\ 
J &    3: 2s$_{1/2}$  &   4: 2p$_{3/2}$  & 6:  3s$_{1/2}$  &  7: 3d$_{3/2}$    &  8: 3p$_{3/2}$  &   8: 3p$_{3/2}$  &   9:  3d$_{5/2}$ &   11: 4s$_{1/2}$  &  12: 4d$_{3/2}$  \\ 
\cline{1-1} \cline{2-10}
   2.0$-$04 &	2.328+2 &  3.172+2 &  1.462+3 &  1.201+3 &  1.756+3 &  7.185+2 &  2.673+3 &  5.089+3 &  4.691+3  \\
   4.0$-$04 &	2.692+2 &  3.463+2 &  1.670+3 &  1.324+3 &  2.026+3 &  7.708+2 &  3.122+3 &  5.759+3 &  5.747+3  \\
   1.0$-$03 &	3.120+2 &  4.039+2 &  1.927+3 &  1.654+3 &  2.549+3 &  8.435+2 &  3.753+3 &  6.637+3 &  7.238+3  \\
   2.0$-$03 &	3.438+2 &  4.674+2 &  2.131+3 &  1.919+3 &  2.979+3 &  8.862+2 &  4.204+3 &  7.308+3 &  8.314+3  \\
   4.0$-$03 &	3.854+2 &  5.397+2 &  2.335+3 &  2.172+3 &  3.384+3 &  9.386+2 &  4.642+3 &  7.976+3 &  9.383+3  \\
   1.0$-$02 &	4.316+2 &  6.697+2 &  2.598+3 &  2.573+3 &  3.896+3 &  1.012+3 &  5.246+3 &  8.855+3 &  1.078+4  \\
   2.0$-$02 &	4.559+2 &  6.971+2 &  2.800+3 &  2.748+3 &  4.311+3 &  1.067+3 &  5.693+3 &  9.517+3 &  1.185+4  \\
   4.0$-$02 &	4.877+2 &  7.758+2 &  3.000+3 &  3.000+3 &  4.711+3 &  1.114+3 &  6.140+3 &  1.018+4 &  1.292+4  \\
   1.0$-$01 &	5.313+2 &  8.852+2 &  3.259+3 &  3.368+3 &  5.233+3 &  1.174+3 &  6.735+3 &  1.106+4 &  1.432+4  \\
   2.0$-$01 &	5.736+2 &  9.148+2 &  3.459+3 &  3.578+3 &  5.632+3 &  1.222+3 &  7.247+3 &  1.172+4 &  1.538+4  \\
   4.0$-$01 &	6.063+2 &  9.805+2 &  3.655+3 &  3.824+3 &  6.024+3 &  1.270+3 &  7.619+3 &  1.238+4 &  1.642+4  \\
   1.0$+$00 &	6.471+2 &  1.080+3 &  3.909+3 &  4.125+3 &  6.529+3 &  1.331+3 &  8.161+3 &  1.324+4 &  1.778+4  \\
   2.0$+$00 &	6.742+2 &  1.133+3 &  4.081+3 &  4.368+3 &  6.875+3 &  1.350+3 &  8.523+3 &  1.385+4 &  1.872+4  \\
   4.0$+$00 &	6.891+2 &  1.180+3 &  4.236+3 &  4.587+3 &  7.184+3 &  1.381+3 &  8.910+3 &  1.437+4 &  1.998+4  \\
   1.0$+$01 &	7.165+2 &  1.235+3 &  4.412+3 &  4.714+3 &  7.534+3 &  1.444+3 &  9.206+3 &  1.496+4 &  2.053+4  \\
   2.0$+$01 &	7.376+2 &  1.260+3 &  4.541+3 &  4.863+3 &  7.796+3 &  1.448+3 &  9.477+3 &  1.541+4 &  2.114+4  \\
   3.0$+$01 &	7.474+2 &  1.280+3 &  4.599+3 &  4.936+3 &  7.912+3 &  1.462+3 &  9.608+3 &  1.560+4 &  2.145+4  \\
   4.0$+$01 &	7.543+2 &  1.294+3 &  4.641+3 &  4.988+3 &  7.955+3 &  1.473+3 &  9.701+3 &  1.574+4 &  2.167+4  \\
   5.0$+$01 &	7.596+2 &  1.304+3 &  4.673+3 &  5.028+3 &  8.059+3 &  1.481+3 &  9.773+3 &  1.585+4 &  2.184+4  \\
\hline
\end{tabular} 
\end{table*}

\clearpage
\setcounter{table}{1} 
\begin{table*}
\caption{continued. }
\begin{tabular}{ccllllllllll} \hline
$E_i$ (Ryd)   & \multicolumn{8}{c}{Transition}         \\   \cline{2-10}
  I &  11: 4s$_{1/2}$ &  12: 4d$_{3/2}$  & 12: 4d$_{3/2}$   &   13: 4p$_{3/2}$  &    14: 4f$_{5/2}$  & 15: 4d$_{5/2}$ &  17: 5p$_{1/2}$  & 17: 5p$_{1/2}$ &  18:  5s$_{1/2}$  \\  
J & 13: 4p$_{3/2}$  &  13:  4p$_{3/2}$ & 14: 4f$_{5/2}$  &   15: 4d$_{5/2}$  &    15: 4d$_{5/2}$  & 16: 4f$_{7/2}$  &  18: 5s$_{1/2}$  & 19: 5d$_{3/2}$ &  20: 5p$_{3/2}$  \\  
\cline{1-1} \cline{2-10}
   2.0$-$04 &	5.816+3 &  3.030+3 &  7.301+3 &  1.163+4 &  1.341+3 &  1.210+4 &  1.318+4 &  1.338+4 &  1.543+4  \\
   4.0$-$04 &	7.134+3 &  3.242+3 &  8.439+3 &  1.360+4 &  1.426+3 &  1.372+4 &  1.485+4 &  1.632+4 &  1.882+4  \\
   1.0$-$03 &	8.935+3 &  3.523+3 &  9.944+3 &  1.617+4 &  1.527+3 &  1.581+4 &  1.707+4 &  2.023+4 &  2.327+4  \\
   2.0$-$03 &	1.033+4 &  3.736+3 &  1.104+4 &  1.804+4 &  1.608+3 &  1.743+4 &  1.874+4 &  2.328+4 &  2.664+4  \\
   4.0$-$03 &	1.163+4 &  3.952+3 &  1.217+4 &  1.998+4 &  1.688+3 &  1.904+4 &  2.040+4 &  2.619+4 &  3.005+4  \\
   1.0$-$02 &	1.342+4 &  4.232+3 &  1.366+4 &  2.253+4 &  1.790+3 &  2.114+4 &  2.259+4 &  3.006+4 &  3.447+4  \\
   2.0$-$02 &	1.475+4 &  4.449+3 &  1.478+4 &  2.447+4 &  1.870+3 &  2.275+4 &  2.426+4 &  3.297+4 &  3.782+4  \\
   4.0$-$02 &	1.608+4 &  4.658+3 &  1.590+4 &  2.638+4 &  1.946+3 &  2.434+4 &  2.591+4 &  3.589+4 &  4.113+4  \\
   1.0$-$01 &	1.784+4 &  4.937+3 &  1.737+4 &  2.890+4 &  2.050+3 &  2.643+4 &  2.810+4 &  3.973+4 &  4.552+4  \\
   2.0$-$01 &	1.917+4 &  5.149+3 &  1.847+4 &  3.080+4 &  2.128+3 &  2.802+4 &  2.976+4 &  4.260+4 &  4.882+4  \\
   4.0$-$01 &	2.050+4 &  5.358+3 &  1.955+4 &  3.267+4 &  2.196+3 &  2.956+4 &  3.140+4 &  4.548+4 &  5.211+4  \\
   1.0$+$00 &	2.221+4 &  5.616+3 &  2.089+4 &  3.512+4 &  2.273+3 &  3.147+4 &  3.355+4 &  4.924+4 &  5.647+4  \\
   2.0$+$00 &	2.342+4 &  5.789+3 &  2.180+4 &  3.681+4 &  2.324+3 &  3.275+4 &  3.510+4 &  5.188+4 &  5.954+4  \\
   4.0$+$00 &	2.457+4 &  5.943+3 &  2.265+4 &  3.924+4 &  2.391+3 &  3.451+4 &  3.644+4 &  5.456+4 &  6.282+4  \\
   1.0$+$01 &	2.585+4 &  6.117+3 &  2.381+4 &  3.990+4 &  2.437+3 &  3.535+4 &  3.794+4 &  5.675+4 &  6.602+4  \\
   2.0$+$01 &	2.654+4 &  6.249+3 &  2.411+4 &  4.113+4 &  2.475+3 &  3.606+4 &  3.906+4 &  5.867+4 &  6.741+4  \\
   3.0$+$01 &	2.693+4 &  6.311+3 &  2.444+4 &  4.169+4 &  2.498+3 &  3.652+4 &  3.954+4 &  5.952+4 &  6.839+4  \\
   4.0$+$01 &	2.720+4 &  6.355+3 &  2.467+4 &  4.209+4 &  2.515+3 &  3.685+4 &  3.989+4 &  6.013+4 &  6.908+4  \\
   5.0$+$01 &	2.742+4 &  6.390+3 &  2.485+4 &  4.240+4 &  2.528+3 &  3.711+4 &  4.016+4 &  6.059+4 &  6.961+4  \\   
\hline
\end{tabular} 
\end{table*}

\clearpage
\setcounter{table}{1}

\begin{table*}
\caption{continued. } 
\begin{tabular}{cclllllllll} \hline
  $E_i$ (Ryd) & \multicolumn{8}{c}{Transition}         \\   \cline{2-9}
  I &  19: 5d$_{3/2}$  & 19: 5d$_{3/2}$ &  20: 5p$_{3/2}$  & 21: 5f$_{5/2}$ &  21: 5f$_{5/2}$  &  22: 5d$_{5/2}$ &   23: 5g$_{7/2}$ &  24:  5f$_{7/2}$ \\  
 J &  20: 5p$_{3/2}$  & 21: 5f$_{5/2}$ &  22: 5d$_{5/2}$  & 22: 5d$_{5/2}$ &  23: 5g$_{7/2}$ &  24: 5f$_{7/2}$ &   24: 5f$_{7/2}$ &  25: 5g$_{9/2}$  \\      
 \cline{1-1} \cline{2-9}
   2.0$-$04 &  8.346+3 &  2.564+4 &  3.257+4 &  4.626+3 &  2.574+4 &  4.290+4 &  2.184+3 &  3.652+4  \\
   4.0$-$04 &  8.925+3 &  2.992+4 &  3.825+4 &  4.908+3 &  2.886+4 &  4.843+4 &  2.307+3 &  4.064+4  \\
   1.0$-$03 &  9.696+3 &  3.518+4 &  4.505+4 &  5.289+3 &  3.319+4 &  5.613+4 &  2.454+3 &  4.618+4  \\
   2.0$-$03 &  1.028+4 &  3.923+4 &  5.038+4 &  5.579+3 &  3.642+4 &  6.190+4 &  2.574+3 &  5.034+4  \\
   4.0$-$03 &  1.086+4 &  4.326+4 &  5.567+4 &  5.856+3 &  3.965+4 &  6.761+4 &  2.695+3 &  5.449+4  \\
   1.0$-$02 &  1.162+4 &  4.856+4 &  6.261+4 &  6.235+3 &  4.388+4 &  7.515+4 &  2.851+3 &  6.000+4  \\
   2.0$-$02 &  1.221+4 &  5.256+4 &  6.785+4 &  6.530+3 &  4.712+4 &  8.086+4 &  2.964+3 &  6.417+4  \\
   4.0$-$02 &  1.279+4 &  5.656+4 &  7.308+4 &  6.811+3 &  5.027+4 &  8.652+4 &  3.085+3 &  6.830+4  \\
   1.0$-$01 &  1.356+4 &  6.179+4 &  7.996+4 &  7.187+3 &  5.449+4 &  9.407+4 &  3.234+3 &  7.375+4  \\
   2.0$-$01 &  1.415+4 &  6.575+4 &  8.515+4 &  7.468+3 &  5.766+4 &  9.971+4 &  3.341+3 &  7.785+4  \\
   4.0$-$01 &  1.472+4 &  6.970+4 &  9.037+4 &  7.746+3 &  6.073+4 &  1.053+5 &  3.435+3 &  8.184+4  \\
   1.0$+$00 &  1.546+4 &  7.474+4 &  9.715+4 &  8.080+3 &  6.456+4 &  1.126+5 &  3.545+3 &  8.677+4  \\
   2.0$+$00 &  1.595+4 &  7.827+4 &  1.019+5 &  8.306+3 &  6.707+4 &  1.176+5 &  3.619+3 &  9.002+4  \\
   4.0$+$00 &  1.639+4 &  8.126+4 &  1.060+5 &  8.507+3 &  6.963+4 &  1.219+5 &  3.753+3 &  9.284+4  \\
   1.0$+$01 &  1.688+4 &  8.461+4 &  1.106+5 &  8.734+3 &  7.172+4 &  1.266+5 &  3.794+3 &  9.607+4  \\
   2.0$+$01 &  1.725+4 &  8.713+4 &  1.140+5 &  8.905+3 &  7.358+4 &  1.302+5 &  3.837+3 &  9.847+4  \\
   3.0$+$01 &  1.742+4 &  8.830+4 &  1.156+5 &  8.989+3 &  7.452+4 &  1.319+5 &  3.872+3 &  9.968+4  \\
   4.0$+$01 &  1.755+4 &  8.913+4 &  1.167+5 &  9.048+3 &  7.519+4 &  1.331+5 &  3.897+3 &  1.005+5  \\
   5.0$+$01 &  1.764+4 &  8.977+4 &  1.175+5 &  9.094+3 &  7.570+4 &  1.340+5 &  3.916+3 &  1.012+5  \\
\hline
\end{tabular} 
\end{table*}

\setcounter{table}{2}

\begin{table*}
\caption{Effective collision strengths for transitions of atomic hydrogen as function of temperature. $a{\pm}b \equiv a{\times}$10$^{{\pm}b}$.}
\begin{tabular}{ccllllllllll} \hline
 $\log_{10} T_e$ (K) &  \multicolumn{8}{c}{Transition}         \\   \cline{2-10}
I  &  2: 2p$_{1/2}$ &   3:  2s$_{1/2}$  &  5: 3p$_{1/2}$   &  5: 3p$_{1/2}$    &  6: 3s$_{1/2}$  &   7: 3d$_{3/2}$  &   8: 3p$_{3/2}$  &   10: 4p$_{1/2}$ &  10: 4p$_{1/2}$  \\ 
J  &   3: 2s$_{1/2}$  &   4: 2p$_{3/2}$  & 6:  3s$_{1/2}$  &  7: 3d$_{3/2}$    &  8: 3p$_{3/2}$  &   8: 3p$_{3/2}$  &   9:  3d$_{5/2}$ &   11: 4s$_{1/2}$  &  12: 4d$_{3/2}$  \\ 
\cline{1-1} \cline{2-10}
   3.00 &  4.100+2 &  6.121+2 &  2.500+3 &  2.403+3 &  3.713+3 &  9.859+2 &  5.021+3 &  8.526+3  & 1.026+4   \\
   3.33 &  4.447+2 &  6.862+2 &  2.719+3 &  2.672+3 &  4.150+3 &  1.042+3 &  5.513+3 &  9.254+3  & 1.143+4   \\
   3.66 &  4.802+2 &  7.611+2 &  2.938+3 &  2.944+3 &  4.588+3 &  1.096+3 &  6.010+3 &  9.982+3  & 1.260+4   \\
   4.00 &  5.188+2 &  8.327+2 &  3.162+3 &  3.222+3 &  5.037+3 &  1.151+3 &  6.525+3 &  1.073+4  & 1.380+4   \\
   4.33 &  5.567+2 &  9.003+2 &  3.378+3 &  3.486+3 &  5.470+3 &  1.203+3 &  7.011+3 &  1.146+4  & 1.495+4   \\
   4.66 &  5.931+2 &  9.697+2 &  3.590+3 &  3.746+3 &  5.896+3 &  1.253+3 &  7.476+3 &  1.217+4  & 1.609+4   \\
   5.00 &  6.272+2 &  1.040+3 &  3.801+3 &  4.011+3 &  6.316+3 &  1.298+3 &  7.932+3 &  1.289+4  & 1.726+4   \\
   5.33 &  6.556+2 &  1.102+3 &  3.990+3 &  4.250+3 &  6.695+3 &  1.338+3 &  8.345+3 &  1.354+4  & 1.835+4   \\
   5.66 &  6.809+2 &  1.156+3 &  4.164+3 &  4.453+3 &  7.042+3 &  1.377+3 &  8.712+3 &  1.413+4  & 1.930+4   \\
   6.00 &  7.051+2 &  1.204+3 &  4.325+3 &  4.634+3 &  7.366+3 &  1.411+3 &  9.045+3 &  1.468+4  & 2.012+4   \\
   6.33 &  7.268+2 &  1.244+3 &  4.466+3 &  4.792+3 &  7.647+3 &  1.439+3 &  9.337+3 &  1.515+4  & 2.082+4   \\
   6.66 &  7.471+2 &  1.282+3 &  4.592+3 &  4.940+3 &  7.901+3 &  1.466+3 &  9.608+3 &  1.558+4  & 2.146+4   \\
   7.00 &  7.666+2 &  1.320+3 &  4.712+3 &  5.084+3 &  8.142+3 &  1.493+3 &  9.871+3 &  1.598+4  & 2.208+4   \\
     \noalign{\smallskip}
\cline{1-10}
   & 11: 4s$_{1/2}$ &  12: 4d$_{3/2}$  & 12: 4d$_{3/2}$   &   13: 4p$_{3/2}$  &    14: 4f$_{5/2}$  & 15: 4d$_{5/2}$ &  17: 5p$_{1/2}$  & 17: 5p$_{1/2}$ &  18:  5s$_{1/2}$  \\  
    $\log_{10} T_e$ (K)  & 13: 4p$_{3/2}$  &  13:  4p$_{3/2}$ & 14: 4f$_{5/2}$  &   15: 4d$_{5/2}$  &    15: 4d$_{5/2}$  & 16: 4f$_{7/2}$  &  18: 5s$_{1/2}$  & 19: 5d$_{3/2}$ &  20: 5p$_{3/2}$  \\  
\cline{1-1} \cline{2-10}
   3.00 &  1.276+4 &  4.128+3 &  1.311+4 &  2.159+4 &  1.751+3 &  2.036+4 &  2.177+4 &  2.862+4  & 3.282+4   \\
   3.33 &  1.422+4 &  4.361+3 &  1.433+4 &  2.370+4 &  1.837+3 &  2.211+4 &  2.360+4 &  3.182+4  & 3.648+4   \\
   3.66 &  1.568+4 &  4.594+3 &  1.556+4 &  2.580+4 &  1.923+3 &  2.385+4 &  2.541+4 &  3.501+4  & 4.012+4   \\
   4.00 &  1.718+4 &  4.832+3 &  1.681+4 &  2.795+4 &  2.010+3 &  2.564+4 &  2.728+4 &  3.828+4  & 4.387+4   \\
   4.33 &  1.864+4 &  5.062+3 &  1.801+4 &  3.003+4 &  2.091+3 &  2.736+4 &  2.909+4 &  4.145+4  & 4.750+4   \\
   4.66 &  2.007+4 &  5.285+3 &  1.917+4 &  3.208+4 &  2.166+3 &  2.902+4 &  3.088+4 &  4.458+4  & 5.110+4   \\
   5.00 &  2.151+4 &  5.502+3 &  2.031+4 &  3.419+4 &  2.237+3 &  3.069+4 &  3.268+4 &  4.774+4  & 5.477+4   \\
   5.33 &  2.284+4 &  5.695+3 &  2.136+4 &  3.617+4 &  2.301+3 &  3.224+4 &  3.432+4 &  5.062+4  & 5.822+4   \\
   5.66 &  2.406+4 &  5.870+3 &  2.231+4 &  3.786+4 &  2.358+3 &  3.358+4 &  3.581+4 &  5.322+4  & 6.136+4   \\
   6.00 &  2.516+4 &  6.034+3 &  2.315+4 &  3.931+4 &  2.410+3 &  3.471+4 &  3.720+4 &  5.558+4  & 6.410+4   \\
   6.33 &  2.608+4 &  6.177+3 &  2.384+4 &  4.055+4 &  2.457+3 &  3.568+4 &  3.840+4 &  5.760+4  & 6.636+4   \\
   6.66 &  2.691+4 &  6.308+3 &  2.448+4 &  4.170+4 &  2.502+3 &  3.659+4 &  3.948+4 &  5.944+4  & 6.839+4   \\
   7.00 &  2.770+4 &  6.433+3 &  2.512+4 &  4.282+4 &  2.546+3 &  3.749+4 &  4.049+4 &  6.119+4  & 7.035+4   \\
\hline
\end{tabular} 

\end{table*}

\clearpage
\setcounter{table}{2}

\begin{table*}
\caption{continued. }
\begin{tabular}{ccllllllllll} \hline
 $\log_{10} T_e$ (K)  & \multicolumn{8}{c}{Transition}         \\   \cline{2-9}
   I & 19: 5d$_{3/2}$  & 19: 5d$_{3/2}$ &  20: 5p$_{3/2}$  & 21: 5f$_{5/2}$ &  21: 5f$_{5/2}$  &  22: 5d$_{5/2}$ &   23: 5g$_{7/2}$ &  24:  5f$_{7/2}$ \\  
 J & 20: 5p$_{3/2}$  & 21: 5f$_{5/2}$ &  22: 5d$_{5/2}$  & 22: 5d$_{5/2}$ &  23: 5g$_{7/2}$ &  24: 5f$_{7/2}$ &   24: 5f$_{7/2}$ &  25: 5g$_{9/2}$  \\     
\cline{1-1} \cline{2-9}
   3.00 &  1.134+4 &  4.657+4 &  6.001+4 &  6.097+3 &  4.230+4 &  7.234+4 &  2.791+3 &  5.795+4   \\
   3.33 &  1.198+4 &  5.096+4 &  6.576+4 &  6.411+3 &  4.581+4 &  7.859+4 &  2.920+3 &  6.250+4   \\
   3.66 &  1.262+4 &  5.534+4 &  7.149+4 &  6.724+3 &  4.931+4 &  8.483+4 &  3.046+3 &  6.704+4   \\
   4.00 &  1.328+4 &  5.982+4 &  7.738+4 &  7.044+3 &  5.290+4 &  9.124+4 &  3.172+3 &  7.168+4   \\
   4.33 &  1.391+4 &  6.415+4 &  8.309+4 &  7.350+3 &  5.633+4 &  9.743+4 &  3.287+3 &  7.613+4   \\
   4.66 &  1.453+4 &  6.841+4 &  8.874+4 &  7.645+3 &  5.964+4 &  1.035+5 &  3.393+3 &  8.042+4   \\
   5.00 &  1.514+4 &  7.262+4 &  9.436+4 &  7.930+3 &  6.288+4 &  1.095+5 &  3.498+3 &  8.455+4   \\
   5.33 &  1.569+4 &  7.639+4 &  9.943+4 &  8.183+3 &  6.576+4 &  1.149+5 &  3.598+3 &  8.819+4   \\
   5.66 &  1.618+4 &  7.980+4 &  1.040+5 &  8.411+3 &  6.829+4 &  1.198+5 &  3.684+3 &  9.146+4   \\
   6.00 &  1.664+4 &  8.296+4 &  1.084+5 &  8.625+3 &  7.060+4 &  1.243+5 &  3.756+3 &  9.451+4   \\
   6.33 &  1.704+4 &  8.571+4 &  1.120+5 &  8.813+3 &  7.264+4 &  1.282+5 &  3.819+3 &  9.720+4   \\
   6.66 &  1.741+4 &  8.821+4 &  1.154+5 &  8.988+3 &  7.455+4 &  1.318+5 &  3.882+3 &  9.970+4   \\
   7.00 &  1.776+4 &  9.059+4 &  1.186+5 &  9.154+3 &  7.641+4 &  1.352+5 &  3.946+3 &  1.021+5   \\
\hline
\end{tabular} 
\end{table*}

\end{document}